%% file: main.tex
\documentclass[conference]{IEEEtran}
\IEEEoverridecommandlockouts
\usepackage{amsmath,amssymb,amsfonts}

\usepackage{svg}
\usepackage{algorithmic}
\usepackage{graphicx}
\usepackage{textcomp}
\usepackage{xcolor}
\usepackage{enumitem}
\usepackage{multicol}
\usepackage{multirow}
\usepackage{booktabs}
\usepackage{xspace}
\usepackage{hyperref}
\usepackage{kotex}
\def\BibTeX{{\rm B\kern-.05em{\sc i\kern-.025em b}\kern-.08em
    T\kern-.1667em\lower.7ex\hbox{E}\kern-.125emX}}

\newcommand{\expone}{\texttt{BGG-CLS}\xspace}
\newcommand{\exptwo}{\texttt{BGG-LOC}\xspace}

\begin{document}

\title{Enemy Spotted: In-game Gun Sound Dataset for Gunshot Classification and Localization\\
{\footnotesize 
}
}

\author{\IEEEauthorblockN{Junwoo Park, Youngwoo Cho, Gyuhyeon Sim, Hojoon Lee, Jaegul Choo}
\IEEEauthorblockA{\textit{Kim Jaechul Graduate School of AI} \\
\textit{Korea Advanced Institute of Science and Technology (KAIST)}\\
Daejeon, South Korea \\
\{junwoo.park, cyw314, ghsim, joonleesky, jchoo\}@kaist.ac.kr
}
}

\maketitle
%

\begin{abstract}
Recently, deep learning-based methods have drawn huge attention due to their simple yet high performance without domain knowledge in sound classification and localization tasks.
However, a lack of gun sounds in existing datasets has been a major obstacle to implementing a support system to spot criminals from their gunshots by leveraging deep learning models.
Since the occurrence of gunshot is rare and unpredictable, it is impractical to collect gun sounds in the real world. 
As an alternative, gun sounds can be obtained from an FPS game that is designed to mimic real-world warfare. The recent FPS game offers a realistic environment where we can safely collect gunshot data while simulating even dangerous situations.
By exploiting the advantage of the game environment, we construct a gunshot dataset, namely BGG, for the firearm classification and gunshot localization tasks.
The BGG dataset consists of 37 different types of firearms, distances, and directions between the sound source and a receiver.
We carefully verify that the in-game gunshot data has sufficient information to identify the location and type of gunshots by training several sound classification and localization baselines on the BGG dataset.
Afterward, we demonstrate that the accuracy of real-world firearm classification and localization tasks can be enhanced by utilizing the BGG dataset. The BGG dataset is available in \url{https://github.com/junwoopark92/BG-Gun-Sound-Dataset}.
\end{abstract}


\section{Introduction}
The first-person shooter (FPS) game, designed to mimic real-world warfare and combat situations, is a popular game genre.
To defeat enemies in the FPS game, a player under attack must decide whether to strike back or retreat by considering the enemies' position and firearms.
However, human vision may not be able to capture where the enemies are and what firearms they have as the distance between the player and them increases. In addition, when the enemies camouflage themselves, it is also difficult to detect the enemies by human vision.
In these cases, gunshots can be clues for estimating the enemies' state.
For example, an expert FPS gamer can recognize the tiny difference in stereophonic sound from a headphone, and she can roughly guess the position and the firearms of the enemies.
The reason the player can establish strategy based on auditory information is that a game engine can reproduce the characteristics of the sound that varies with distance and direction.
Inspired by the realism of the game, we hypothesize that a prediction model, which localizes enemies and identifies firearms from in-game gunshots, can also be applied to real-world gunshots.
Specifically, a support system that can spot the enemies and determine the type of firearm from gunshots does not only assist the beginners of the game, but also aids soldiers and police officers who track the criminals in the real world.

The problem of spotting enemies and identifying the firearms can be formulated as a sound localization and classification task.
The sound localization task aims at estimating the distance and angle between a sound source and a receiver, and the sound classification task aims at categorizing the sound into predefined classes.
Recently, deep learning has become the standard practice in sound localization~\cite{yiwere2017distance, yiwere2020sound} and classification~\cite{choi2017convolutional, cakir2017convolutional} tasks. 
However, a lack of gunshot samples in existing datasets~\cite{45857, salamon2014dataset} has been a major obstacle to building the support system in surveillance and military service by leveraging the deep learning models.
A real-world gunshot dataset, \texttt{Forensic}~\cite{raponi2020sound}, alleviates this problem by firing and recording in a controlled environment.
Nevertheless, collecting data in the real world is expensive and may not cover the engagement situation due to the dangers of firearms.

As an alternative, the FPS games enable us to easily collect data by simulating engagement situations in which a bullet passes by the side of the face.
By exploiting this advantage of the simulation, we constructed a battleground gunshot (\texttt{BGG}) dataset collected from PlayerUnknown's Battleground\footnote{https://krafton.com/games/battlegrounds} (PUBG).
PUBG is a representative FPS game where 100 players aim to survive by eliminating each other with firearms and other weapons until only one player (or team) remains.
We recorded the gun sounds and labeled the types of firearms, their distance, and angle by varying the enemies' position.

Before applying the in-game gunshots to real-world tasks, we evaluated baselines widely used in sound classification and Transformer~\cite{vaswani2017attention}, which is underexplored in gunshot classification and localization tasks.
As a result, we confirmed that the in-game gunshots sufficiently reflect the characteristics of the real-world sound and the models can learn those characteristics to predict the location enemies and their firearms.
Furthermore, we empirically demonstrated that our in-game gunshot dataset can improve the accuracy of both sound classification and gunshot localization in real-world datasets.

The contributions of our work are summarized as follows:
\begin{itemize}[leftmargin=*]
\item We construct a gunshot dataset, \texttt{BGG}, for gunshot classification and localization by exploiting the advantages of the game environment.

\item We verified that the deep learning models can spot enemies position and identify their firearms from the in-game gunshots by presenting the application deployed in PUBG.

\item We demonstrated that the BGG dataset can be utilized to improve the accuracy of the real-world sound classification and gunshot localization tasks.

\end{itemize}

\section{Related work}
In this work, we deal with two problems, sound classification and localization, which are the tasks of identifying the type of sound and estimating the location of sound sources.
For decades, both tasks have been studied as a crucial component for various domains such as robotics~\cite{5354419,yalta2017sound}, speech recognition~\cite{7497454}, and surveillance system~\cite{6636622}.
Generally, sound localization is more challenging than sound classification since the number of sources, the interaction between sources and surroundings, and the movement of sources should be considered~\cite{grumiaux2021survey} in the case of localization.
The fundamental idea of sound localization is to leverage the difference in arrival time or phase of signals obtained from the microphone array.
Currently, well-known sound source localization methods include multiple signal classification (MUSIC)~\cite{1143830} and generalized cross-correlation with phase transform (GCC-PHAT)~\cite{1162830}.
Recently, data-driven approaches, (\textit{i.e.,} sound classification and localization based on deep learning), have been actively studied~\cite{grumiaux2021survey}.
Most of the deep learning-based approaches employ deep convolutional neural networks (DCNNs)~\cite{yalta2017sound} or convolutional recurrent neural networks (CRNNs)~\cite{cakir2017convolutional}, and each method varies in the architectural details, the input features, and the output type.

An application of sound classification and localization is gunshot classification and localization, which focuses on gunshots as sound sources, and several systems for detecting gunshots have been proposed for achieving social safety~\cite{hansen2021gunshot,6637700,9141621}.
A major obstacle in the gunshot classification and localization task is the data acquisition problem.
Even though there are several datasets containing gunshot samples, the number of gunshot-related audio samples is fewer than that of the other classes.
For instance, the number of gun sound samples in \texttt{AudioSet}~\cite{45857} and \texttt{UrbanSound8k}~\cite{salamon2014dataset} account for only 0.2\% and 4.2\%, respectively, since the occurrence of gunshots is rare and unpredictable.
Therefore, predictions from deep learning models trained on the existing datasets are inaccurate for gunshots compared to other types of sound.
Singh et al.~\cite{baliram2021data} reported that sound source localization models could not distinguish gunshots and gunshot-like audio events (\textit{e.g.,} plastic bag bursting) on the \texttt{UrbanSound8k}.
To address this issue, Raponi et al.~\cite{raponi2020sound} proposed a gunshot-dedicated dataset, \texttt{Forensic}, a collection of gunshots recorded in a controlled environment.
However, the application of the \texttt{Forensic} is limited to a few specific scenarios because the \texttt{Forensic} does not reflect the actual engagement situation.
As an alternative, several data-efficient methods were proposed to deal with the class imbalance and data deficiency problems~\cite{lim2017rare,shimada2020metric}.
Shimada et al.~\cite{shimada2020metric} introduced a few-shot sound classification method that classifies rare classes (\textit{e.g.,} gunshot, glassbreak, and babycry) by utilizing a metric-learning algorithm. 

\section{BGG Dataset}

\input{tables/dataset_stat.tex}
Unlike other types of sound, it is difficult to collect gunshots in real world since firing a gun is a rare and dangerous event.
One may collect gunshots in a shooting range, but it is expensive, and the sounds could be reproduced in limited situations.
In this case, a game environment is a reasonable option for acquiring gunshots.
For this reason, we constructed the BGG dataset, and this section covers the detailed description of our dataset.
\subsection{Dataset Descriptions}
We recorded the gun sounds by changing the type and position of guns to diversify distances and angles in the PUBG environment.
As shown in Table~\ref{tab:stats}, the BGG dataset consists of 2,195 samples with 37 different types of guns and five directions, including a silence in which there is no gunfire, but noises exist.
The distance from the firearms ranged from 0 meters to 600 meters.
Audio was recorded in stereo (\textit{i.e.,} two-channel audio), and each sample contains various environmental noises (\textit{e.g.,} water splashing, walking, and bullet friction).

\input{figures/exp1_dist.tex}
\input{figures/exp2_dist.tex}
We constructed the two datasets from the collected sounds, namely \expone and \exptwo dataset.
\expone was made for training and evaluating firearm classification models.
Figure~\ref{fig:exp1_dist} shows the class distribution of \expone.
The class denoted by \textit{No gun} means that there is no gunfire and it only contains the sound of noise.
We defined this class to prevent models from predicting the gun types in the absence of gunshots.
Using \exptwo dataset, we can train and evaluate models that estimate the distance and direction of a gunshot.
\exptwo dataset contains 1,024 sound samples categorized into three guns and \textit{No gun}, and each sample is labeled with six distances (from 0 meters to 600 meters) and five directions (\textit{front, back, left, right} and \textit{center}).
The distribution of each class is shown in Figure~\ref{fig:exp2_dist}.
Our dataset contains longer distances and more gun types than the existing datasets~\cite{45857, raponi2020sound}.
To the best of our knowledge, there is no dataset for sound localization that addresses gunshots measured up to 600 meters.

\subsection{Analysis of the Gun Sound}
Figure~\ref{fig:waveform} visualizes the audio samples in a two-channel waveform according to the gun type, distance, and direction. We checked whether the sound obtained from the game reflects the characteristics of the real-world sound.
The volume of the gunshot decreases as the distance between an enemy and the player increases.
At this time, the waveform of bullet friction with the air around the player (see the blunt shape in 600 meters in Figure~\ref{fig:waveform}) is more prominent than the waveform of gunshot (see the sharp shape in 0 meters).
This phenomenon is observed when the player is far away from the enemy since the bullet friction sound is closer to the player than the location of the gunshot. In addition, the volume difference between the two channels appeared depending on the direction the player is facing and the location of the enemy's gunshot.
Meanwhile, in the case of automatic weapon, the firing rate varies depending on the gun type. This difference can be seen in the spacing of the gunshots on the waveform.
For example, firing rate of \textit{M416} is 0.085 seconds, which is faster than \textit{AKM} of which rate is 0.100 seconds.
From the aforementioned observations, we assume that classifying firearms and localizing the enemy are possible from the gunshots obtained in the game environment.

\section{Method}
\label{sec:method}
This section formulates the sound classification and localization tasks, describes model architectures, and explains the training details.


\subsection{Preliminaries}
\input{figures/waveforms.tex} 

As a preliminary, we first define the notation of our dataset. 
Our \texttt{BGG} dataset is given as $\mathcal{D} = \{(x_n, y_n^g, y_n^s, y_n^d)\}_{n=1}^N$ where $N$ indicates the number of instances.
Here, $x_n \in \mathbb{R}^{T\times C}$ indicates a sound input where $C$ denotes the number of channels and $T$ represents the number of time steps.
There exist three different types of labels; the one-hot representations of the sound type $y_n^g$, the discretized distance $y_n^s$, and the discretized direction $y_n^d$.
Given a parameterized model $f_{\theta}$, the goal of sound classification is to train the model which accurately predicts the sound type label $y_n^g$ given $x_n$.
Similarly, the goal of sound localization is to estimate the distance $y_n^s$ and the direction $y_n^d$ labels given $x_n$.

\subsection{Model Architecture}

Here, we describe the architectural details of our parameterized model $f_{\theta}$. 
The model is divided into two components (i) feature extractor ${F}_{\theta}$ and (ii) classifier $C_{\theta}$.
For simplicity, we omit the instance subscript $n$ throughout this section.
First, an audio signal represented in a waveform $x \in \mathbb{R}^{C\times T}$ is converted to a spectrogram by Short-time Fourier transform (STFT) operation, which is widely used to detect how the frequency changes over time. 
Second, the processed spectrogram is passed on to the feature extractor $F_{\theta}$ to obtain a dense representation of the input sound.
This module aggregates the spectrogram over the timestep. 
\begin{equation}
    z = {F}_{\theta}(\text{\texttt{STFT}}(x))
\end{equation}
where $z \in \mathbb{R}^{D}$ is aggregated along the time-axis, and $D$ denotes the hidden dimension size.

Finally, the extracted representation $z$ is passed onto the classification module $C_{\theta}$. 
The classification module consists of two fully-connected layers with a ReLU nonlinearity. 
Thereby, we obtain the predicted gunshot class probability $p^g$, distance class probability $p^s$, and  and direction probability $p^d$.
\begin{equation}
    p^g, p^s, p^d = {C}_{\theta}^{g}(z), {C}_{\theta}^{s}(z), {C}_{\theta}^{d}(z) 
\end{equation}



\subsection{Training}

For the gunshot classification dataset, \expone, 
model parameters $\theta$ are optimized to maximize the probability $p^g$ of a gun type label $y^g$ for a given gunshot.
Cross-entropy loss is used to to maximize the log-likelihood.
\begin{equation}
    \mathcal{L}_{cls}=- y^g \log p^g
\end{equation}

For the gunshot localization dataset, \exptwo, the model parameters $\theta$ are trained to maximize the probability $p^s$ and the probability $p^d$ of the ground truth distance label $y^s$ and direction label $y^d$, respectively.
As a form of multitask learning, we utilize the cross-entropy loss to optimize the log-likelihood.
\begin{equation}
    \mathcal{L}_{loc}= - (y^s \log p^s + y^d \log p^d)
\end{equation}



\vspace{2mm}

\section{Experiments}
In this section, we describe the results of firearm classification and gunshot localization on \texttt{BGG} dataset.
We evaluated three widely used models and two Transformer-based models~\cite{vaswani2017attention}, which are underexplored in firearm classification and gunshot localization tasks, to analyze the quality of our dataset.

\subsection{Experimental Setting}
We employed three sound classification methods: DCNNs~\cite{yalta2017sound}, LSTM~\cite{lezhenin2019urban}, and CRNNs~\cite{cakir2017convolutional}.
DCNNs use stacked convolution layers where each layer is sequentially constructed with a 1D convolution operation, batch normalization, and nonlinearity functions.
Convolutional layers can learn the temporal dependency of a spectrogram by sliding the filter of fixed size along with the time axis.
LSTM is also widely used for learning temporal dependencies in sequential inputs while alleviating the vanishing gradient problem of RNNs via long-term and short-term memories.
LSTM can be a suitable option for audio data whose time length increases according to a sampling rate compared to RNNs.
We used bidirectional LSTM (bi-LSTM) in the experiment since the sound classification task has no restrictions on precedence in setting problems.
CNNs are advantageous for learning local patterns, but have limitations in discovering long-distance temporal dependencies since they require more layers than RNNs.
LSTMs are powerful in long-term dependency with fewer parameters than CNNs.
Combining two methods, CRNNs showed considerable improvements in sound classification tasks.
We additionally added a Transformer-based models.
Transformer~\cite{vaswani2017attention} has shown successful results by replacing CNNs and RNNs in the natural language processing, but has been relatively underexplored in firearm classification.
The self-attention mechanism in Transformer is known to successfully learn dependencies between words in long sentences.
Similar to CRNNs, we tested Transformer and CNN-Transformer.

We adopt two widely used classification metrics, accuracy (acc.) and F1 score, to measure the classification and localization performance.
Accuracy is calculated by dividing the number of correctly predicted samples by the total number of samples.
F1 score is measured by computing the harmonic mean of the precision and recall, and this measurement can also consider the class imbalance.

\vspace{1mm}
\subsection{Experiments Results}
\input{tables/bgg_ssc.tex}
We verified that deep learning models can predict firearms from in-game gunshots.
Table~\ref{bgg_ssc} presents the performance of the five models trained on \expone dataset.
The best accuracy and F1 score are reached by CNN-Transformer that uses convolutional layers as feature extractor and then integrates the extracted features via Transformer.
The DCNN and CRNN that show competitive performance with the best model also utilize convolutional layers as a feature extractor.
On the other hand, Bi-LSTM and Transformer showed poor accuracy and F1 score.
These results empirically indicate that local temporal patterns in the spectrogram are necessary to classify firearms. 

\input{tables/bgg_ssl.tex}
For the gunshot localization task, we trained and evaluated the models on \exptwo to predict the distance and direction between the enemies and the game player.
Since we formulated the localization task as the classification problem, the feature extractor of the models which learn discriminative features from the spectrogram is identical to the firearm classification models.
The only difference is that there are two classifiers to predict the direction and distance.
As shown in Table~\ref{bgg_ssl}, we confirmed that the overall performance of the DCNN, CRNN, and CNN-Transformer was higher than the Bi-LSTM and Transformer, showing the superiority of convolutional layers as a feature extractor.
Note that the accuracy and F1 score of the models trained on \exptwo are higher than that of the models trained on \expone, because the number of firearm classes of \expone is smaller than that of \exptwo.

Unlike when predicting a firearm, Bi-LSTM and Transformer showed competitive performance with the DCNN, and the CRNN outperformed the DCNN when predicting distance.
Figure~\ref{fig:waveform} provides the possible explanation for this result.
The enemies prefer aiming over continuous fire as the distance from the observer increases.
Therefore, there are differences in the number of gunshots and the interval between them during the same time, depending on the distance.
Although recurrent unit and self-attention mechanism can learn these temporal patterns, the max pooling layer of DCNN can ignore the differences because it only extracts large values along the time axis.
Thus, we can reliably say that the temporal pattern is important as well as the difference in sound volume depending on the distance.

When predicting the direction of gunshots, DCNN and CNN-Transformer showed better performance than the other three methods.
In the sight of the human auditory sense, the left and right ears correspond to two waveform channels and receive different audio signals depending on the direction of the sound due to the distance between the ears.
In Figure~\ref{fig:waveform}, we observed the volume difference between the two channels.
A max pooling after convolutional layers effectively learns these differences to predict direction.
Finally, we confirmed that CNN-Transformer was the best method when comprehensively evaluating the three tasks, followed by the DCNN model.
As a result, we designed a gunshot detection system in the game.
The following section describes the qualitative results of the application.


\subsection{Qualitative Result of the Application}
\input{figures/demo.tex}
This section demonstrates the qualitative results when our support system is deployed in the PUBG.
We recorded the in-game video and extracted the audio signals from it.
Given the extracted audio signal, our application estimates the type, distance and direction of gunshots according to time. The test scenario is as follows: 

\begin{enumerate}[label=(\alph*)]
\item For the first five seconds, a player is vigilant about his surroundings.
\item An enemy at 100 meters to the right starts to fire an \textit{AKM}.
\item The player recognizes the attack and begins to strike back with an \textit{M416} after facing the enemy.
\item Finally, by judging that there is no chance of winning, the player runs away in the opposite direction of the enemy. 
\end{enumerate}
As shown in Figure~\ref{fig:demos}, our application accurately detected the alternating \textit{AKM} and \textit{M416} gunshots in this scenario.
Furthermore, in predicting the direction of gunshot, we confirmed that the application could detect the transition when the player turns his head to face the enemy (\textit{i.e.,} from right to front). 
However, there is room for improvement in the distance estimation.
According to the scenario, since the gunfires occur alternately at 100 meters and 0 meters, only the probability of 0 meters and 100 meters should be increased, but the probability of 50 meters increased.

\subsection{Improving Performance in Real-World Benchmarks}
A sufficient amount of data is necessary to achieve successful results using a deep learning model.
However, the number of gunshot samples is smaller than the other sound samples in the existing datasets since gunshots occur less frequently than other sounds.
Moreover, even if data is collected by firing guns in the real world, it is challenging to reproduce various situations and obtain detailed labeling due to the danger of firearms.
Beyond the entertainment role, FPS games enable easily collecting data by simulating even dangerous situations in which a bullet passes by the side of the face.
We show that the performance of gunshot-related tasks can be improved in real-world benchmarks through simple approaches using our BGG dataset.
We reported the mean values measured in 10 experiments while mixing the training, validation, and test datasets.

\input{tables/bgg2urban.tex}
The UrbanSound8K (Urban) dataset includes ten different classes of sounds, of which only one is the gunshot class.
Compared to 8,358 samples belonging to classes other than gunshots, the number of samples included in the gunshot class is the smallest at 374, resulting in the lowest classification accuracy.
We compared models trained on the urban dataset additionally containing gunshots from the BGG as the samples of the urban gunshot class with models trained by simply augmenting the gunshot samples in the existing urban by applying random cropping and speed-changing methods.
For fair comparison, the number of added data was set to the same number for both models.
Table~\ref{bgg2urban} shows that the accuracy and F1 score are improved with a significant gap when the BGG is used.
The large gap in Gun acc. means that the performance improvement was attributed to the model's ability to distinguish gun sounds by adding the BGG.

\input{tables/bgg2foren.tex}

To validate in-game data in gunshot classification and localization tasks, we trained and evaluated the baseline models on the Forensic (Foren) dataset, consisting of 2,241 samples categorized into 18 different types of firearms, eight distances, and 13 directions.
Unlike the Urban dataset, which also contains non-gunshots, the Foren dataset contains only gunshots.
After training models that predict firearms, distance, and direction on the BGG dataset, we use the parameters of the feature extractor as the initial parameters of the same network classifying firearms, distance, and direction into predefined classes of the Foren.
We compared the models that were pretrained on the BGG and those that were only trained on the Foren from scratch.
As shown in Table~\ref{bgg2foren}, the models trained on the BGG outperformed the models trained from scratch in accuracy and F1 score.
Although the BGG dataset and the Foren dataset contain different firearms, the parameters pretrained with the in-game gunshots in the BGG dataset strengthen the ability of the model to classify the real-world gunshots into firearms from the Foren dataset.
In the case of predicting direction and distance, improvement in accuracy when the BGG is used is not significant because the range of distance and direction covered by the Foren and the BGG is different. Nevertheless, except for three cases, all performance has increased. Thus, we can say that it can be used to create a support system in the real world by using in-game data.

\section{Conclusion}
Recognizing enemies from a gunshot has substantial benefits to prepare for the threat in FPS games and the real world.
In the real world, the support system that identifies and locates gunshots plays a vital role in protecting social safety from the misuse of guns.
However, due to the rarity and unpredictable occurrence of gunshots in the real world, the number of gun sound samples is insufficient to build deep learning-based gunshot classification and localization models.
As an alternative, by recording gunshots in an FPS game, we constructed a gunshot dataset for the firearm classification and localization tasks.
We validated the practicality of the in-game gunshot data by analyzing the deep-learning models trained on our BGG dataset.
Moreover, we demonstrated that the gun sound dataset collected from the game improves real-world gunshot classification and localization accuracy.

\section*{Acknowledgements}
This work was supported by Institute of Information \& communications Technology Planning \& Evaluation (IITP) grant funded by the Korea government (MSIT) (No.2019-0-00075, Artificial Intelligence Graduate School Program (KAIST)) and the National Research Foundation of Korea (NRF) grant funded by the Korean government (MSIT) (No. NRF-2022R1A2B5B02001913).

\bibliography{reference}

\clearpage

\end{document}

%% file: tables/dataset_stat.tex
\begin{table}
\label{tab:stats}
\caption{Basic statistics of BGG dataset. The values in the parentheses are the number of samples and classes that are not gunshots.}
\centering
\begin{tabular}{lcccc}
\toprule
& BGG & Foren & Urban \\
\midrule
\# of Gun sound samples & 2,195 (57) &  2,241 & 374 (8,358) \\
Range of audio length (sec.) & [3,8] & [1.5,2.5] & [0.05,4]\\
Sampling rate of audio (KHz) & 44.1 & 96 & [8,192] \\
\# of Gun categories & 37 (1) & 18 & 1 (9) \\
\# of Directions from source & 5 (1) & 13 & - \\
\# of Distances from source & 6 (1) & 8 & - \\
Range of distance (m) & [0,600] & [0,150] & -\\
\bottomrule
\end{tabular}
\end{table}

%% file: figures/exp1_dist.tex
\begin{figure*}[h!]
\begin{center}
  \includegraphics[width=\textwidth]{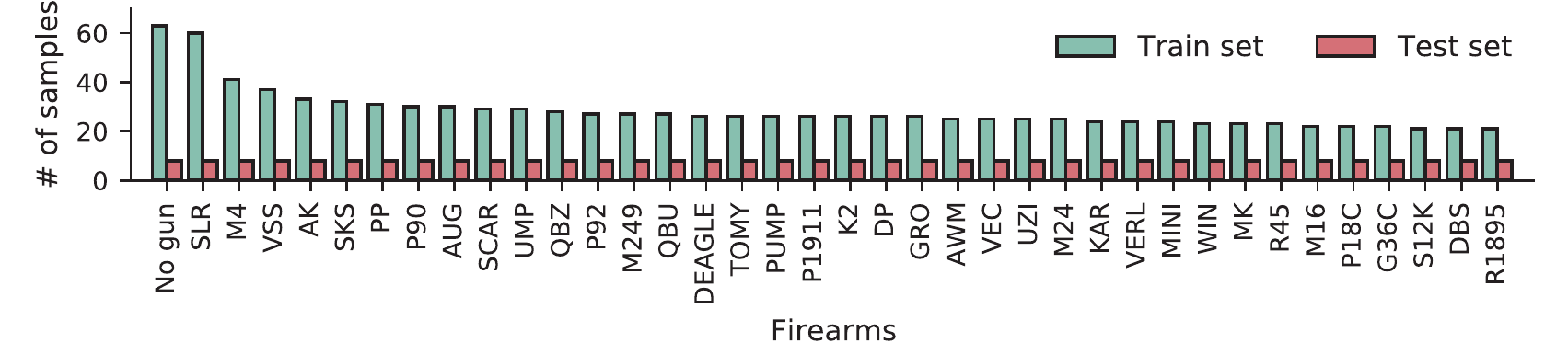}
\end{center}
\vspace*{-0.2cm}
   \caption{The distribution of firearms in the \expone dataset is plotted. The blue and red bars indicate the number of train and test samples, respectively.}
\label{fig:exp1_dist}
\end{figure*}

%% file: figures/exp2_dist.tex
\begin{figure*}[h!]
\begin{center}
  \includegraphics[width=\textwidth]{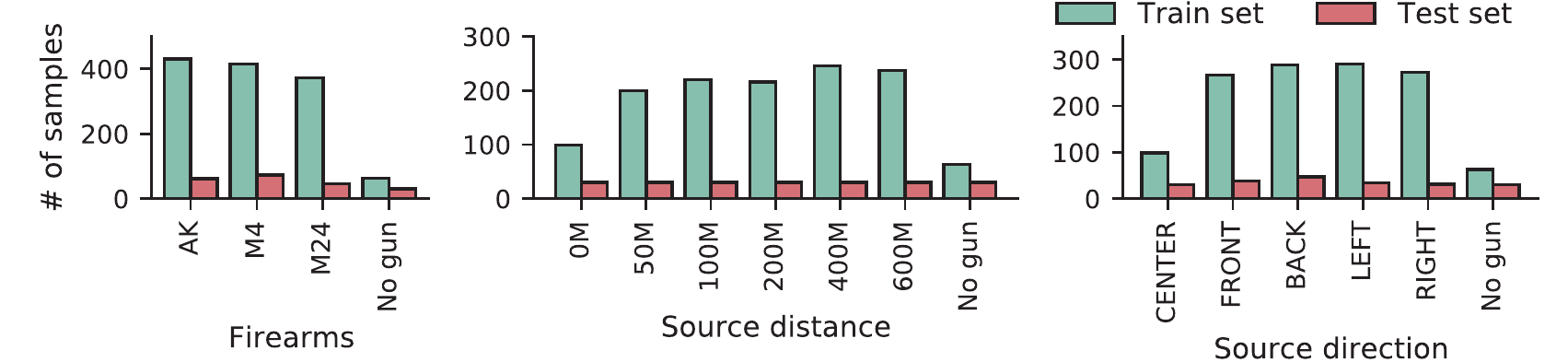}
\end{center}
\vspace*{-0.2cm}
   \caption{The distributions of firearms, source distance, and source direction in the \exptwo dataset are plotted. The blue and red bars indicate the number of train and test samples, respectively.}
\label{fig:exp2_dist}
\end{figure*}

%% file: figures/waveforms.tex
\begin{figure*}[h!]
\begin{center}
  \includegraphics[width=1.0\linewidth]{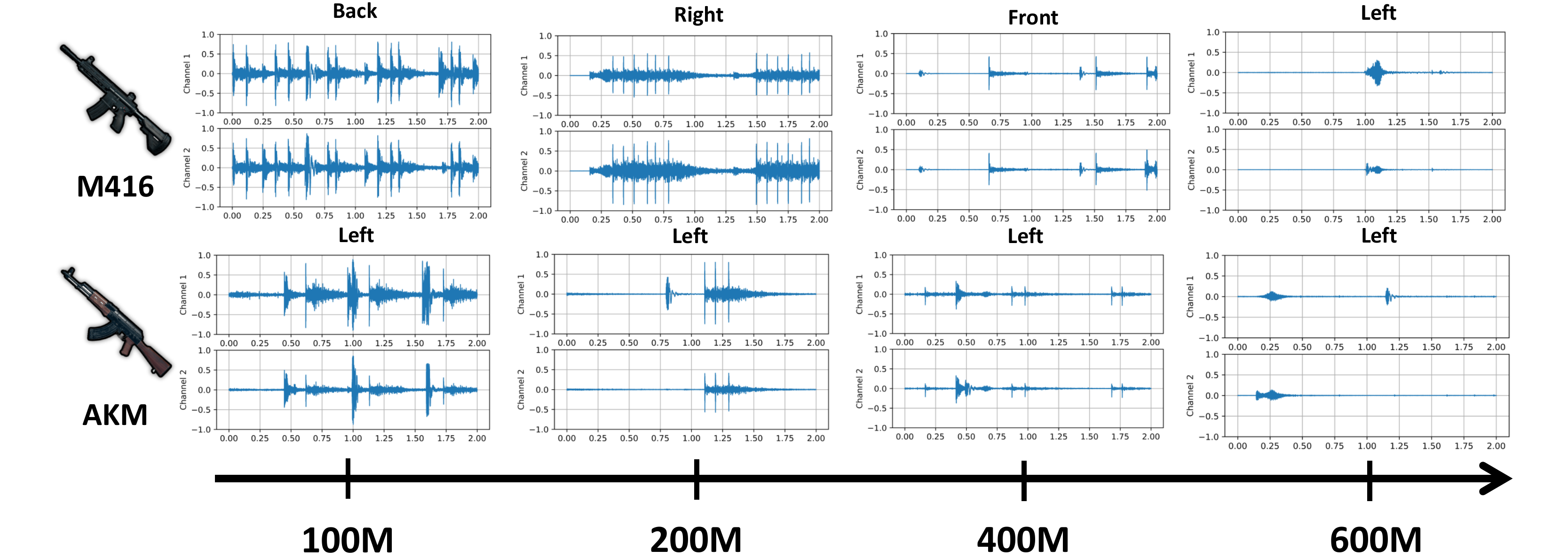}
\end{center}
\vspace*{-0.2cm}
   \caption{Sample audio samples are visualized in the waveform. All samples are recorded in stereo. For each sound sample, the left and the right channel audio samples are plotted at the top and the bottom, respectively.}
\label{fig:waveform}
\end{figure*}

%% file: tables/bgg_ssc.tex


\begin{table}[h]
\centering
\caption{Sound classification results on \expone dataset that consists of 37 gun classes evaluated with accuracy and F1 score.}
\label{bgg_ssc}
\begin{tabular}{@{}lcccc@{}}
\toprule
\multicolumn{1}{c}{\multirow{2}{*}{Method}} & \multicolumn{2}{c}{Validation}    & \multicolumn{2}{c}{Test}          \\
\multicolumn{1}{c}{}                        & Acc.            & F1              & Acc.            & F1              \\ \midrule
Bi-LSTM                                     & 0.6627          & 0.5695          & 0.6068          & 0.5739          \\
DCNN                                         & 0.8753          & 0.8381          & 0.8503          & 0.8315          \\
CRNN                                        & 0.8445          & 0.8304          & 0.8279          & 0.8009          \\
Transformer                                 & 0.6348          & 0.5571          & 0.5918          & 0.5642          \\
CNN-Transformer                             & \textbf{0.9384} & \textbf{0.9196} & \textbf{0.9360} & \textbf{0.9290} \\ \bottomrule
\end{tabular}
\end{table}

%% file: tables/bgg_ssl.tex
\begin{table*}
\caption{Gunshot localization results over \exptwo dataset that consists of three gun classes, six directions, and seven distances.}
\vspace{2mm}
\label{bgg_ssl}
\centering
\begin{tabular}{l|c|cccccc}
\toprule
\multirow{2}{*}{\textbf{Method}} & \multirow{2}{*}{\textbf{Rank}}& \multicolumn{2}{c}{ \textbf{Firearm}} & \multicolumn{2}{c}{\textbf{Distance}} & \multicolumn{2}{c}{\textbf{Direction}} \\
& &   Acc & F1 & Acc & F1 & Acc & F1 \\
\midrule
Bi-LSTM  & 5 & 0.8717 $\pm$ 0.0210 & 0.8776  $\pm$ 0.0190 & 0.8593 $\pm$ 0.0189 & 0.8596 $\pm$ 0.0169 & 0.7143 $\pm$ 0.0274 &0.7257 $\pm$ 0.0274 \\
DCNN & 2 & \underline{0.9507} $\pm$ 0.0107 & \textbf{0.9424} $\pm$ 0.0115 & 0.8668 $\pm$ 0.0129 & 0.8695 $\pm$ 0.0121 & \underline{0.8792} $\pm$ 0.0065  & \textbf{0.9099} $\pm$ 0.0112  \\
CRNN & 3 & 0.9339 $\pm$ 0.0112 & 0.9292 $\pm$ 0.0112 & \underline{0.8901} $\pm$ 0.0150 & \underline{0.8912} $\pm$ 0.0149 & 0.7900 $\pm$ 0.0738 & 0.7789 $\pm$ 0.0681 \\
Transformer & 4 & 0.9011 $\pm$ 0.0131 & 0.9015 $\pm$ 0.0124 & 0.8327 $\pm$ 0.0162 &0.8307 $\pm$ 0.0159 & 0.7815 $\pm$ 0.0214 & 0.8079 $\pm$ 0.0184 \\
CNN-Transformer & 1 & \textbf{0.9529} $\pm$ 0.0054 & \underline{0.9386} $\pm$ 0.0075 &\textbf{0.9150} $\pm$ 0.0165 & \textbf{0.9156} $\pm$  0.0159 & \textbf{0.9313} $\pm$  0.0085 & \underline{0.9003} $\pm$ 0.0132 \\
\bottomrule
\end{tabular}
\end{table*}

%% file: figures/demo.tex
\begin{figure*}[h!]
\begin{center}
\includegraphics[width=\textwidth]{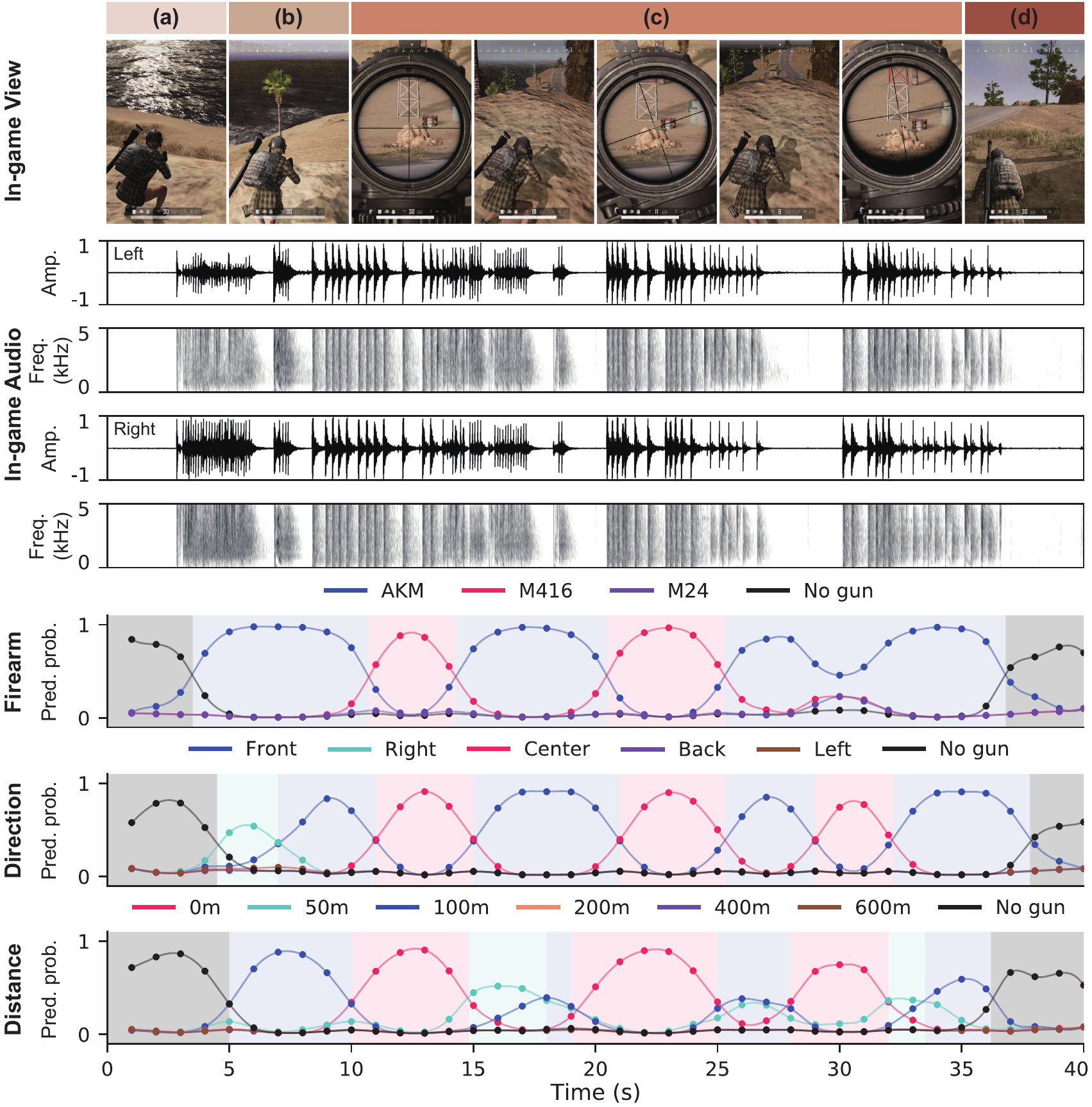}
\end{center}
\vspace*{-0.3cm}
\caption{Demonstration of our gunshot classification and localization application on the PUBG environment.
(\textbf{a}), (\textbf{b}), (\textbf{c}), and (\textbf{d}) refer to each phase in the confrontation with the enemy.
Along with in-game screenshots, a sound sample recorded from the game is plotted in the raw waveform and the spectrogram.
Three rows at the bottom visualize the prediction results from firearm classification and gunshot localization model.
The visualizations are highlighted based on the prediction probabilities.
At the beginning and the end of the engagement, no gunshot occurs in the environment (\textbf{a} and \textbf{d}).
The player turns right to find the enemy when the enemy starts to attack the player (\textbf{b}), and the predicted direction changes from right to front (highlights changes from \textit{sky-blue} to \textit{dark-blue}).
During engagement (\textbf{c}), the player and the enemy fire their firearms alternatively; therefore, we can see all predictions aligned (highlights changes between \textit{dark-blue} and \textit{red}).
}
\label{fig:demos}

\end{figure*}

%% file: tables/bgg2urban.tex
\begin{table}[h]
\centering
\caption{The result of sound classification over Urban dataset evaluated with accuracy and F1 score. We report the mean of the values obtained from 10 cases.}
\label{bgg2urban}
\begin{tabular}{@{}llcc@{}}
\toprule
\multicolumn{1}{c}{\multirow{2}{*}{\textbf{Method}}} & \multicolumn{1}{c}{} & \multicolumn{2}{c}{Urban only → Urban + BGG} \\
\multicolumn{1}{c}{}                        & \multicolumn{1}{c}{} & Valid                 & Test                 \\ \midrule
\multirow{3}{*}{DCNN}                        & Acc.                 & 0.7556 → \textbf{0.7748}       & 0.7370 → \textbf{0.7736}      \\
                                            & F1                   & 0.7542 → \textbf{0.7718}       & 0.7206 → \textbf{0.7650}      \\
                                            & Gun acc.             & 0.5747 → \textbf{0.6272}               & 0.2661 → \textbf{0.4708}      \\ \cmidrule(l){2-4} 
\multirow{3}{*}{CRNN}                       & Acc.                 & 0.7456 → \textbf{0.7616}       & 0.7386 → \textbf{0.7680}      \\
                                            & F1                   & 0.7442 → \textbf{0.7615}       &0.7158 → \textbf{0.7567}      \\
                                            & Gun acc.             & 0.4600 → \textbf{0.5051}                    & 0.1998 → \textbf{0.3841}      \\ \cmidrule(l){2-4} 
\multirow{3}{*}{CNN-Transformer}            & Acc.                 & 0.7726 → \textbf{0.7963}       & 0.7693 → \textbf{0.8006}      \\
                                            & F1                   & 0.7696 → \textbf{0.7946}       & 0.7495 → \textbf{0.7951}      \\
                                            & Gun acc.             & 0.5661 → \textbf{0.6534}                    & 0.2709 → \textbf{0.5249}      \\ \bottomrule
\end{tabular}
\end{table}

%% file: tables/bgg2foren.tex

\begin{table}[h]
\centering
\caption{The results of gunshot classification and localization over Foren dataset evaluated with accuracy and F1 score. We report the mean of the values obtained from 10 cases.}
\label{bgg2foren}
\begin{tabular}{@{}llcc@{}}
\toprule
\multicolumn{1}{c}{\multirow{2}{*}{\textbf{Firearm}}} & \multicolumn{1}{c}{\multirow{2}{*}{}} & \multicolumn{2}{c}{Foren only → Foren + BGG} \\
\multicolumn{1}{c}{}                        & \multicolumn{1}{c}{}                  & Valid                 & Test                 \\ \midrule
\multirow{2}{*}{DCNN}                        & Acc.                                  & 0.6696 → \textbf{0.6968}       & 0.6832 → \textbf{0.7016}      \\
                                            & F1                                    & 0.6154 → \textbf{0.6518}       & 0.6200 → \textbf{0.6509}      \\ \cmidrule(l){2-4} 
\multirow{2}{*}{CRNN}                       & Acc.                                  & 0.6181 → \textbf{0.6692}       & 0.6096 → \textbf{0.6616}      \\
                                            & F1                                    & 0.5856 → \textbf{0.6386}       & 0.5726 → \textbf{0.6236}      \\ \cmidrule(l){2-4} 
\multirow{2}{*}{CNN-Transformer}            & Acc.                                  & 0.6741 → \textbf{0.7260}       & 0.6798 → \textbf{0.7410}      \\
                                            & F1                                    & 0.6362 → \textbf{0.7082}      & 0.6346 → \textbf{0.7126}      \\ \bottomrule
\end{tabular}

\vspace*{0.3cm}

\begin{tabular}{@{}llcc@{}}
\toprule
\multicolumn{1}{c}{\multirow{2}{*}{\textbf{Distance}}} & \multicolumn{1}{c}{\multirow{2}{*}{}} & \multicolumn{2}{c}{Foren only → Foren + BGG} \\
\multicolumn{1}{c}{}                        & \multicolumn{1}{c}{}                  & Valid                 & Test                 \\ \midrule
\multirow{2}{*}{DCNN}                       & Acc.                                  & 0.6351 → \textbf{0.6611}       & 0.6691 → \textbf{0.6941}      \\
                                            & F1                                    & 0.5053 → \textbf{0.5319}       & 0.5148 → \textbf{0.5500}      \\ \cmidrule(l){2-4} 
\multirow{2}{*}{CRNN}                       & Acc.                                  & 0.7501 → \textbf{0.7836}       & 0.7763 → \textbf{0.7946}      \\
                                            & F1                                    & 0.6732 → \textbf{0.6995}       & 0.6988 → \textbf{0.7049}      \\ \cmidrule(l){2-4} 
\multirow{2}{*}{CNN-Transformer}            & Acc.                                  & 0.8429 → \textbf{0.8526}       & 0.8566 → \textbf{0.8641}      \\
                                            & F1                                    & 0.8192 → \textbf{0.8538}       & 0.8323 → \textbf{0.8558}      \\ \bottomrule
\end{tabular}

\vspace*{0.3cm}

\begin{tabular}{@{}llcc@{}}
\toprule
\multicolumn{1}{c}{\multirow{2}{*}{\textbf{Direction}}} & \multicolumn{1}{c}{\multirow{2}{*}{}} & \multicolumn{2}{c}{Foren only → Foren + BGG} \\
\multicolumn{1}{c}{}                        & \multicolumn{1}{c}{}                  & Valid                 & Test                 \\ \midrule
\multirow{2}{*}{DCNN}                       & Acc.                                  & 0.7163 → \textbf{0.7560}       & 0.7820 → \textbf{0.8016}      \\
                                            & F1                                    & 0.5970 → \textbf{0.6319}      & 0.6266 → \textbf{0.6722}      \\ \cmidrule(l){2-4} 
\multirow{2}{*}{CRNN}                       & Acc.                                  & 0.7402 → \textbf{0.7439}       & 0.8426 → 0.8033      \\
                                            & F1                                    & 0.6898 → \textbf{0.6999}      & 0.7280 → 0.7188      \\ \cmidrule(l){2-4} 
\multirow{2}{*}{CNN-Transformer}            & Acc.                                  & 0.8628 → \textbf{0.8713}       & 0.8607 → \textbf{0.8827}      \\
                                            & F1                                    & 0.8141 → \textbf{0.8376}       & 0.8421 → 0.8420      \\ \bottomrule
\end{tabular}

\end{table}